\documentclass[a4paper]{article}
\usepackage[numbers]{natbib}
\usepackage{graphicx}
\usepackage{dcolumn}
\usepackage{bm}
\usepackage{threeparttable}
\usepackage{float}
\usepackage{authblk}

\title{Light speed variation from active galactic nuclei\footnote{Published in Science Bulletin~65~(2020)~262-266.}}

\author[a]{Hao Li}
\author[a,b,c,d]{Bo-Qiang Ma\footnote{Corresponding author. E-mail address: mabq@pku.edu.cn}}

\affil[a]{School of Physics and State Key Laboratory of Nuclear Physics and
	Technology, Peking University, Beijing 100871,
	China}

\affil[b]{Collaborative Innovation Center of Quantum Matter, Beijing 100871, China}

\affil[c]{Center for High Energy Physics, Peking University, Beijing 100871, China}
\affil[d]{Center for History and Philosophy of Science, Peking University, Beijing 100871, China}

\date{}

\begin{document}

\maketitle

\begin{abstract}
	Recent studies on the high-energy photons from gamma-ray bursts~(GRBs) suggested a light speed variation $v(E)=c(1-E/E_{\mathrm{LV}})$ with $E_\mathrm{LV}=3.6\times 10^{17}$~GeV. We check this speed variation from previous observations on light curves of three active galactic nuclei~(AGNs), namely Markarian 421 (Mrk 421), Markarian 501 (Mrk 501) and PKS 2155-304. We show that several phenomena related to the light curves of these AGNs can serve as the supports for the light speed variation determined from GRBs.
\end{abstract}

\section*{Introduction}


According to Einstein's relativity, the speed of light is a constant in free space. However, it is speculated from quantum gravity that the light speed may receive
a correction in powers of $E/E_\mathrm{Pl}$, where $E$ is the energy of the photon and $E_\mathrm{Pl}= \sqrt{\hbar c^5 / G} \approx 1.22 \times 10^{19}$~GeV is the Planck energy. Therefore the light speed
might be energy dependent due to the Lorentz invariance violation (see, e.g., Refs.~\cite{Amelino-Camelia1997a,AmelinoCamelia:2008qg,Ellis:2008gg,Li:2009tt,Xiao:2009xe,Shao:2010wk}). Such speed variation is very tiny and can hardly be detected on the earth. It was suggested by Amelino-Camelia {\it et al.}~\cite{AmelinoCamelia:1997gz}
that photons from gamma-ray bursts~(GRBs) can be used to detect the light speed variation.

As the energy of the photon is extremely small compared to the Planck scale,  the light speed can be written in a Taylor expansion~\cite{Xiao:2009xe} with higher-order terms dropped as
\begin{equation}
v (E) = c\left[1-s_n\frac{n+1}{2}(\frac{E}{E_{\mathrm{LV}, n}})^{n}\right],
\end{equation}
where $n = 1$ or $2$ corresponds to linear or quadratic energy dependence of light speed respectively, $s_n=+1$ or $-1$ corresponds to subluminal or superluminal cases respectively, and $E_{\mathrm{LV},n}$ is the Lorentz violation (LV) scale to be determined by observed data.
As a result, such speed modification can cause an arrival time lag between two photons emitted simultaneously from the source but with different energies. With the cosmological expansion taken into account, the time lag is calculated as~\cite{Jacob:2008bw,Ellis:2002in}
\begin{equation}
\Delta  t_\mathrm{LV} = s_{n}\frac{1+{n}}{2H_0}\frac{E^{n}_{h}-E^{n}_{l}}{E^{n}_{\mathrm{LV}, n}}\int_0^{z}\frac{(1+{z}^\prime)^{n} d{z}^\prime}{\sqrt{\Omega_\mathrm{m}(1+{z}^\prime)^3+\Omega_\mathrm{\Lambda}}},
\label{eq}
\end{equation}
where $E_{h}$ and $E_{l}$ are the observed energies of the high- and low-energy photons, $z$ is the redshift of the source, $H_0$ is the present day Hubble expansion rate taken as $67.3\pm 1.2$ $\mathrm{kms^{-1}Mpc^{-1}}$~\cite{Agashe:2014kda}, and $\Omega_\mathrm{m} = 0.315^{+0.016}_{-0.017}$ and $\Omega_\mathrm{\Lambda} = 0.685^{+0.017}_{-0.016}$ are the pressureless matter density and the dark energy density of the $\mathrm{\Lambda{}CDM}$ Universe~\cite{Agashe:2014kda}.

However, photons are normally emitted at different time from the source, so the intrinsic time lag $\Delta t_\mathrm{in}$ must be taken into account and the observed time lag between photons should be written as~\cite{Ellis:2005wr}
\begin{equation}
\Delta t_\mathrm{obs} = t_\mathrm{high} - t_\mathrm{low} = \Delta t_\mathrm{LV} + (1+{z})\Delta t_\mathrm{in}. \label{4}
\end{equation}

Based on the above scenario, researches on gamma-ray bursts have been carried out in Refs.~\cite{Shao:2009bv,Zhang:2014wpb,Xu:2016zxi,Xu:2016zsa,Xu:2018ien,Amelino-Camelia:2016ohi,Liu:2018qrg} with the finding of a regularity that a number of high-energy photons from different GRBs fall on a same mainline to indicate a light speed variation with the first-order Lorentz violation scale determined as $3.60 \times 10^{17}$~GeV~\cite{Xu:2016zxi,Xu:2016zsa,Xu:2018ien}. In the following, we use $E_\mathrm{LV}^\mathrm{grb}$ to denote this Lorentz violation scale as it is determined from photons of gamma-ray bursts.

Besides GRBs, other astrophysical objects also emit high-energy photons, from which we can study the light speed variation as a supplement to test $E_\mathrm{LV}^\mathrm{grb}$ from GRBs.
One kind of these objects, active galactic nuclei~(AGNs), on which we focus in this paper, are the most luminous persistent sources of electromagnetic radiation
at the center of galaxies.
The purpose of this paper is to exam the light speed variation from the light curves of three AGNs. We show that several observations reported previously
can serve as the supports for the light speed variation with the determined $E_\mathrm{LV}^\mathrm{grb}$.


We focus on three active galactic nuclei~(AGNs) with small redshifts, namely Markarian 421 (Mrk 421), Markarian 501 (Mrk 501) and PKS 2155-304.
The energies of photons from AGNs can reach up to 10~TeV while photons from GRBs observed by Fermi telescope usually carry energies no more than one hundred GeV. As a result, AGNs can also serve as ideal astrophysical objects to study the light speed variation though they are not far from the earth than GRBs. Previous studies~\cite{Biller:1998hg,Albert:2007qk,Albert:2007zd,Aharonian:2008kz} proposed lower limits on the Lorentz violation scale $E_\mathrm{LV}$ without a definite conclusion for the light speed variation. We discuss their conclusions in detail in the following.

Now that we have the LV scale $E_\mathrm{LV}^\mathrm{grb}$ determined from GRBs, we carry out this research in a different way compared to earlier researches. We assume that some high- and low-energy photons, like the ones corresponding to the light curve peaks, were emitted at the same time, and we use Eq.(\ref{4}) ($\Delta t_\mathrm{in}$ now is 0) to calculate the time difference predicted by the determined $E_\mathrm{LV}^\mathrm{grb}$ and compare this time difference with the observed time lag to verify the light speed variation.

\section*{Markarian 421}

Markarian 421 (Mrk 421) is a blazar located around 397 million light-years to 434 million light-years from the earth. It was observed by the Whipple Observatory $\gamma$-ray telescope located in Arizona, and the telescope detects the \v Cerenkov light generated by electromagnetic cascades resulting from the interaction of high-energy $\gamma$ rays in the atmosphere. In an earlier paper~\cite{Biller:1998hg}, the Lorentz violation effect is studied with the data of the rapid flare with a redshift $z = 0.031$ of Mrk 421 on 15 May 1996. The data are separated into two groups, less than 1~TeV and more than 2~TeV, according to their energies. Both groups are binned in intervals of 280~s duration. The light curves of both groups show no difference between the peaks of them, so a time lag which is no more than 280~s was recommended.
With $E_\mathrm{LV}^\mathrm{grb}$, we can predict a time delay of 38.6~s per TeV energy difference. While in Ref.~\cite{Biller:1998hg} with data binned in intervals of 280 seconds, only a time delay larger than 280~s can be detectable, so in the absence of source effects, there could be no distinguishable time delay unless the mean energy difference between the two bands is larger than 7.25~TeV, which leads to a time delay of $7.25\times38.6\approx280$~s. Considering that the energy spectra referred to in Refs.~\cite{Krennrich:1998ana, Biller:1998hg} was approximated by an $\sim E^{-2.5}$ power law between energies 300~GeV and 10~TeV, we are able to estimate the mean energy difference as about 6~TeV, which is less than 7~TeV, so it is reasonable that there is no observation of time lag as reported in Ref.~\cite{Biller:1998hg}. The discussion above indicates that the possible light speed variation scale from photons of Mrk 421 is compatible with that of GRBs, i.e., with $E_\mathrm{LV}^\mathrm{grb}$ in Refs.~\cite{Xu:2016zxi, Xu:2016zsa, Xu:2018ien}.

\section*{Markarian 501}

A flare of Markarian 501 (Mrk 501) observed by Major Atmospheric Gamma-ray Imaging \v Cerenkov (MAGIC) telescope during the night on 9 July 2005~\cite{Albert:2007zd} has been used
to constrain $E_\mathrm{LV}$ in a previous study~\cite{Albert:2007qk}.
The redshift of Mrk 501 is 0.034, similar to that of Mrk 421. In Ref.~\cite{Albert:2007zd}, a $4 \pm 1$ minute time delay between the peak of the light curve of energy band 0.15-0.25~TeV and that of energy band 1.2-10~TeV was reported.
Modified synchrotron-self-Compton mechanisms were mentioned as the possible origin of this time delay~\cite{Albert:2007zd,Albert:2007qk}. In Ref.~\cite{Albert:2007qk}, a lower limit of $2.1 \times 10^{17}~\mathrm{GeV}$ on $E_\mathrm{LV}$ is recommended from analysis of the data.

We indicate in the following that the above observation of a $4 \pm 1$ minute time delay can serve as an evidence to support the determined $E_\mathrm{LV}^\mathrm{grb}$ for the light speed variation~\cite{Xu:2016zxi,Xu:2016zsa,Xu:2018ien}.

\begin{figure}
	\centering
	\includegraphics[width = 8.6cm]{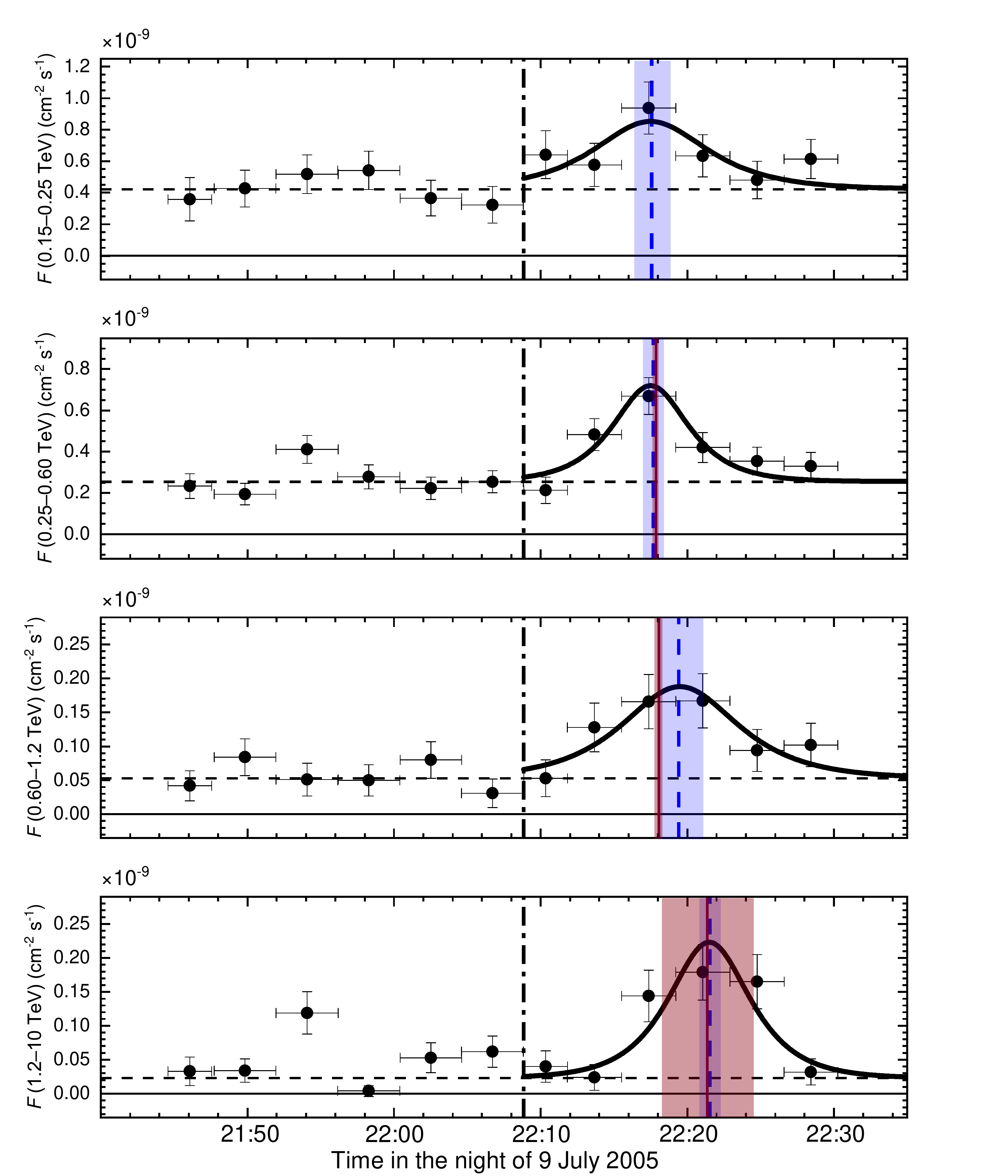}
	\caption{Light curves (LC) binned in 4 minutes for the flare of Mrk 501 in the night on 9 July 2005, see Fig.7 in Ref.~\cite{Albert:2007zd}. The LCs are separated in different energy bands, i.e., from top to bottom, 0.15-0.25, 0.25-0.60, 0.60-1.2 and 1.2-10~TeV. The vertical error bar denotes 1$\sigma$ statistical uncertainties. The vertical black dot-dashed line separates the data into pre-burst and in-burst emission. The vertical blue dashed lines show the positions of the peaks of these four light curves, and the blue regions show the possible peak positions of the light curves with errors taken into account. The vertical red solid lines show the predicted peak positions determined by $\hat{T}$s in Table~\ref{table2}, and the red regions show the possible positions of the predicted peaks determined by $\tilde{T}$s in Table~\ref{table2}. The horizontal black dashed lines represent the averages of the pre-burst emissions. The in-burst emissions were fitted with the flare model described by Eq.~\ref{flaremodel} under the assumption $c = d$. The fitting parameters are listed in Table~\ref{table1}.	\label{figure1}}
\end{figure}

The flare amplitude, duration and its rise/fall times can be combined by~\cite{Albert:2007zd}
\begin{equation}
F(t) = a + \frac{b}{2^{-\frac{t-t_0}{c}}+2^{\frac{t-t_0}{d}}},
\label{flaremodel}
\end{equation}
where $a$ is assumed to be a constant at the time of the flare, $t_0$ is set to the time corresponding to the highest point in the light curve, and $b$, $c$, $d$ need to be determined by fitting the data. In addition, an assumption is made that $c$ equals $d$, or in other words, rise time equals fall time. All light curves of different energy bands were fitted with the flare model described by Eq.~(\ref{flaremodel}), with the resulting parameters listed in Table~\ref{table1}. Meanwhile, in Table~\ref{table1}, we also list the median energy $E_m$ of each energy band which will be useful later. The combined fitting gives $\chi ^2 /\mathrm{NDF} = 14.0/12 (\mathrm{P} = 0.3)$~\cite{Albert:2007zd}, which means that the assumption of $c = d$ is compatible with the fitting. The light curves and fitting curves are shown in Fig.~\ref{figure1}, and instantly we can recognize the time difference between the peaks of different fitting curves.

For convenience, we denote the four energy bands of 0.15-0.25~TeV, 0.25-0.6~TeV, 0.6-1.2~TeV, and 1.2-10~TeV as Band 1, 2, 3, 4 respectively, and denote the observed time difference between the peak of Band $i$ and that of Band $j$ as $T(i,j)$, {where $i,j = \mathrm{1, 2, 3, 4}$}. In addition, with the known $E_\mathrm{LV}^\mathrm{grb}$, Eq.~(\ref{4}) and the assumption that $\Delta t_\mathrm{in} = 0$, we can derive the time delay between the peaks of two different energy bands predicted with $E_\mathrm{LV}^\mathrm{grb}$ in two different ways: one is a time delay range $\tilde{T}(i,j)$ with its upper boundary calculated with the upper limit of the higher energy band and the lower limit of the lower energy band, and the lower boundary in reverse; another one is a precise time delay $\hat{T}(i,j)$ calculated with the median energies of the two bands. Then we list all the observed time differences $T(i,j)$, the time delay ranges $\tilde{T}(i,j)$ and the precise time delays $\hat{T}(i,j)$ predicted with $E_\mathrm{LV}^\mathrm{grb}$ in Table~\ref{table2}.

We immediately find that all time delay ranges represented by $\tilde{T}$ are consistent with the observed time delays when errors are taken into account, indicating that the light speed variation with $E_\mathrm{LV}^\mathrm{grb}$  can explain the time delays between the light curve peaks of different energy bands. Besides, because of the shortage of data of high energy photons, a relatively large band has been taken from 1.2~TeV to 10~TeV. Therefore it is difficult to determine the specific energy the peak corresponds to and we take the median energies of all energy bands for simplicity and consistency. As a result, all $\hat{T}$s in Table~\ref{table2} match the observed time delays $T$ in Table~\ref{table2}. According to those results, the light speed variation with $E_\mathrm{LV}^\mathrm{grb}$ can explain the time delays between different peaks.


\begin{table*}
	\caption{Flare model parameters of the combined fit for the flare of Mrk501 night on 9 July 2005 with $c = d$~\cite{Albert:2007zd} and the median energy of each band}
	\centering
	\begin{threeparttable}
		
		 \resizebox{\textwidth}{12mm}{
		\begin{tabular}{cccccc}\\
			\hline\hline
			Energy Band & a\tnote{a} & b & c & $t_0 - t_0^{\mathrm{LC}, 0.15-0.25}$~TeV{}\tnote{b} & $E_{m}$\tnote{c}\\
			(TeV) & ($10^{-10}$ photons $\mathrm{cm}^{-2}$~$\mathrm{s}^{-1}$) & ($10^{-10}$ photons $\mathrm{cm}^{-2}$~$\mathrm{s}^{-1}$) & (s) & (s) & (TeV)\\
			\hline
			0.15-0.25 & 4.23 $\pm$ 0.49 & 8.6 $\pm$ 3.7 & 143 $\pm$ 92 & 0 $\pm$ 68 & 0.20 \\
			0.25-0.60 & 2.55 $\pm$ 0.24 & 9.3 $\pm$ 2.5 & 95 $\pm$ 28 & 7 $\pm$ 36 & 0.425 \\
			0.6-1.2 & 0.53 $\pm$ 0.10 & 2.7 $\pm$ 0.9 & 146 $\pm$ 56 & 111 $\pm$ 91 & 0.9 \\
			1.2-10 & 0.23 $\pm$ 0.06 & 4.0 $\pm$ 0.9 & 103 $\pm$ 19 & 239 $\pm$ 40 & 5.6 \\
			\hline\hline
		\end{tabular}}
		\begin{tablenotes}
			\item[a] Integrated pre-burst emission flux.\\
			\item[b] $t_0^{\mathrm{LC}, 0.15-0.25}$ is the $t_0$ for the light curve of energy band 0.15-0.25~TeV.\\
			\item[c] The median energy of each energy band.
		\end{tablenotes}
	\end{threeparttable}	
	\label{table1}
\end{table*}

\begin{table}
	\caption{Time delays from observed data and the light speed variation with $E_\mathrm{LV}^\mathrm{grb}$}
	\centering
	\begin{tabular}{cccc}
		\hline
		Selected Energy Bands & $T$ & $\tilde{T}$ & $\hat{T}$ \\
		& (s) & (s) & (s) \\
		\hline
		1 and 2 & 7 $\pm$ 77 & 0-19.7 & 9.8\\
		1 and 3 & 111 $\pm$ 114 & 15.3-45.9 & 30.6 \\
		1 and 4 & 239 $\pm$ 79 & 41.5-430.3 & 235.9 \\
		2 and 3 & 104 $\pm$ 98 & 0-41.5 & 20.7 \\
		2 and 4 & 232 $\pm$ 54 & 26.2-425.9 & 226.0 \\
		3 and 4 & 128 $\pm$ 99 & 0-410.6 & 205.3 \\
		\hline
	\end{tabular}
	\label{table2}
\end{table}

Reversely, we can use the observed time delay of 239~s between Bands 1 and 4 to calculate $E_\mathrm{LV}$ from Eq.~(\ref{eq}) and get $E_\mathrm{LV}=3.55\times 10^{17}~\mathrm{GeV}$.
We then carry out a linear fitting of Eq.~(\ref{eq}) with all observed time delays $T(i,j)$ listed in Table~\ref{table2} to calculate $E_\mathrm{LV}$ from AGNs and get $E^\mathrm{agn}_\mathrm{LV}=3.68^{+0.46}_{-0.37} \times 10^{17}~\mathrm{GeV}$
with Pearson's $r = 0.9704$.
It is straightforward to see that $E_\mathrm{LV}^\mathrm{agn}$ with errors matches $E_\mathrm{LV}^\mathrm{grb}$ well. The surprising consistency between $E_\mathrm{LV}^\mathrm{agn}$ from AGNs and $E_\mathrm{LV}^\mathrm{grb}$ from GRBs can be considered as a strong support for the light speed variation
revealed in Refs.~\cite{Xu:2016zxi,Xu:2016zsa,Xu:2018ien}

\section*{PKS 2155-304}%

\begin{figure}
	\centering
	\includegraphics[width=8.6cm]{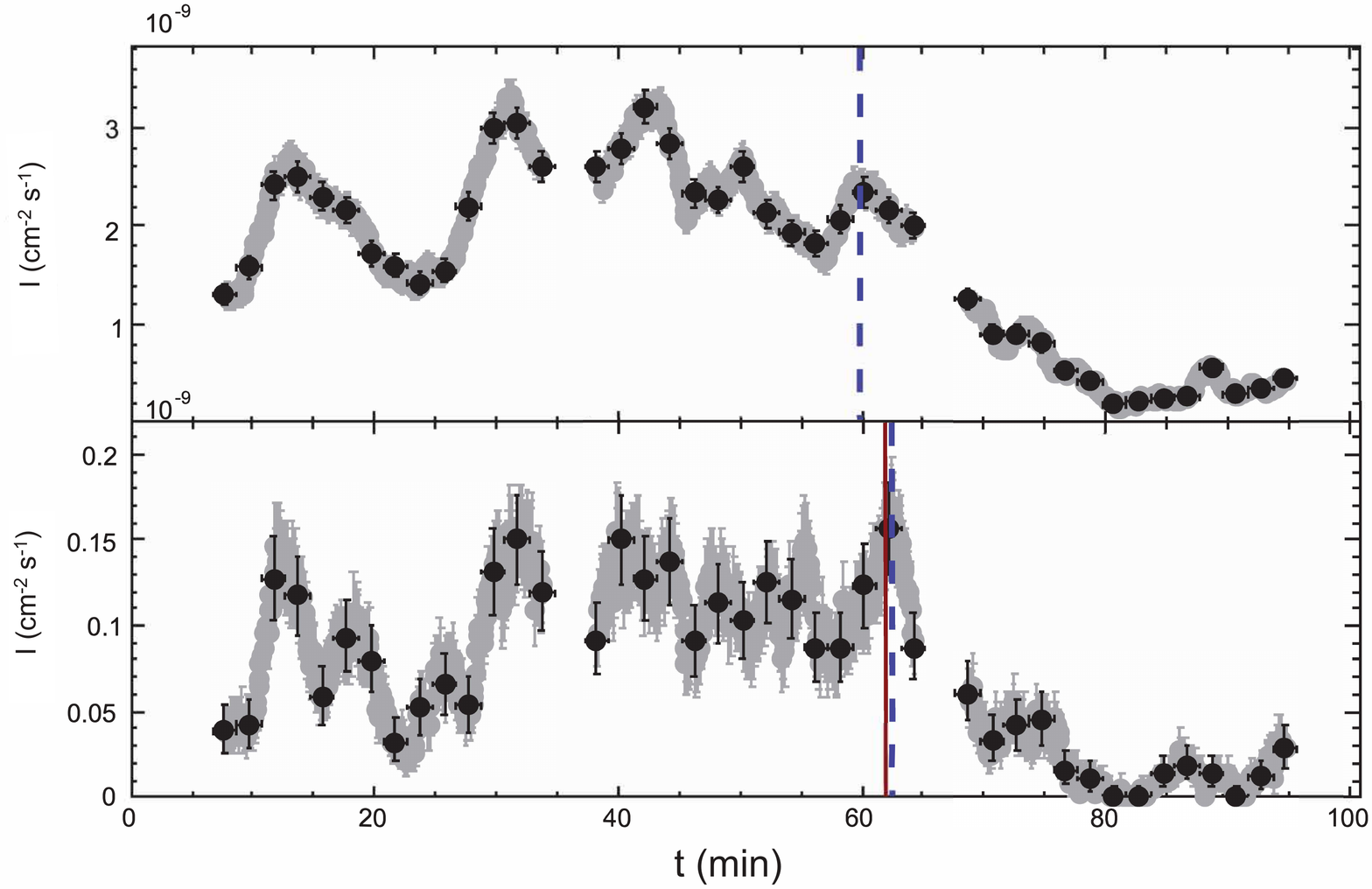}
	\centering
	\caption{Light curves (black points) of energy bands 200-800~GeV (upper panel) and $>$800~GeV (lower panel), binned in 2-minute time intervals, see Fig.1 of Ref.~\cite{Aharonian:2008kz}. For the oversampled light curves (gray points) the 2 min. bins are shifted in units of 5 sec. The vertical blue dashed lines show the positions of the last peaks of the two bands, and the vertical red solid line shows the predicted peak position with light speed variation from GRBs. In addition, the zero time point is set to MJD 53944.02.}
	\label{figure2}
\end{figure}

The object PKS 2155-304, which is located at the redshift
$z = 0.116$, is an active galactic nucleus observed by the High Energy Stereoscopic System (H.E.S.S.) on 28 July 2006~\cite{Falomo:1993dv}. In a previous paper~\cite{Aharonian:2008kz}, time delays between light curves of different energy bands were studied in order to set a limit on Lorentz violation scale. In that paper~\cite{Aharonian:2008kz}, a method which can determine the time lag between two light curves with the modified cross correlation function (MCCF) was adopted~\cite{Li:2004kp}. Two energy bands of 200-800~GeV and more than 800~GeV, with mean energy difference of 1.0~TeV, were studied. A time lag about 20~s with errors around 30~s was given from analyzing the oversampled light curves of the two bands as shown in Fig.~\ref{figure2} with the MCCF method, leading to a lower bound of $E_\mathrm{LV}> 7.2\times10^{17}$~GeV~\cite{Aharonian:2008kz}, which seems incompatible
with $E_\mathrm{LV}^\mathrm{grb}=3.6\times 10^{17}$~GeV from GRBs~\cite{Xu:2016zxi,Xu:2016zsa,Xu:2018ien}.

However, with Eq.~(\ref{4}) and the assumption that $\Delta t_\mathrm{in} = 0$, the time lag caused by the light speed variation with $E_\mathrm{LV}^\mathrm{grb}$ is predicted to be 136~s, which is larger than the upper limit 73~s given in Ref.~\cite{Aharonian:2008kz}.
But when we check the light curves in Fig.~\ref{figure2}, there is no clear one-to-one correspondence between peaks for the first several peaks of the two light curves,
indicating the inadequacy to compare the light curves between the two bands in the whole time range.
We find that if we focus only on the last peaks of both the light curves, a time difference around 2 minutes, which is compatible with the outcome 136~s predicted with $E_\mathrm{LV}^\mathrm{grb}$, appears. This single correlation between the last peaks of the two bands can be considered as a signal for the light speed variation with $E_\mathrm{LV}^\mathrm{grb}$~\cite{Xu:2016zxi,Xu:2016zsa,Xu:2018ien}.

\vspace{5mm}

In this paper, we just simply ignore the intrinsic time $\Delta t_\mathrm{in}$.
As a focus of future researches, we suggest that it is important to adopt a method that can distinguish intrinsic time caused by source effects from the light speed variation induced time delays.  What is more, the assumption that source effects can be ignored may cover up the possible light speed variation effect just as we can see from PKS 2155-304. In order to reveal the possible source effects of AGNs from the light speed variation effect, we may combine several AGNs with similar features but different redshifts. To be specific, we can go back to Eq.~(\ref{4}) and use the same method in Refs.~\cite{Xu:2016zxi, Xu:2016zsa} to fit the parameters $E_\mathrm{LV}^\mathrm{agn}$ and $\Delta t_\mathrm{in}$ with more data from AGNs with different redshifts. We expect that there are similar redshift independent source effects of AGNs of similar features in light curves when more data will be available.
We recommend more data of AGNs from available observatories such as LHAASO~\cite{Cao:2010zz,Zhen:2014zpa,Bai:2019khm} (which begins to operate right now) should be combined to reveal the possible light speed variation effect from other intrinsic source effects.

In summary, in this work we studied the light curves of three AGNs with very high energy flares. We indicate that the light speed variation with determined $E_\mathrm{LV}^\mathrm{grb}$ in Refs.~\cite{Xu:2016zxi,Xu:2016zsa,Xu:2018ien} can account for the time delay of Mrk 501 reported in Ref.~\cite{Albert:2007zd}. The surprising consistency between $E^\mathrm{agn}_\mathrm{LV}=3.68^{+0.46}_{-0.37} \times 10^{17}~\mathrm{GeV}$ from Mrk 501 and $E_\mathrm{LV}^\mathrm{grb}=(3.60\pm0.26)\times 10^{17}$~GeV from GRBs can be considered as a robust support for the light speed variation
revealed in Refs.~\cite{Xu:2016zxi,Xu:2016zsa,Xu:2018ien}.
Meanwhile we illustrate that the time delay caused by the light speed variation for Mrk 421 is 38.6~s per TeV energy difference, too small to be observed as the data are binned in intervals of 280~s duration. As a result, the light speed variation scale from Mrk 421 is compatible with that of GRBs, i.e., with $E_\mathrm{LV}^\mathrm{grb}$ in Refs.~\cite{Xu:2016zxi, Xu:2016zsa, Xu:2018ien}.
We also find that a simple correlation between the last peaks of the light curves of the 200-800~GeV band and the $>$800~GeV band of PKS 2155-304
can be considered as a signal for the light speed variation.
Solely for each observation, one cannot draw a definite conclusion on the light speed variation as was done in previous studies. However, if we combine these AGN observations with
the light speed variation determined from GRBs in Refs.~\cite{Xu:2016zxi,Xu:2016zsa,Xu:2018ien}, we can consider these phenomena as the supports for the light speed variation at a scale $E_\mathrm{LV}=3.6\times 10^{17}$~GeV.



\section*{Acknowledgments}
This work is supported by National Natural Science Foundation of China (Grant No.~11475006) and Huabao Student Research Collaborative Innovation Fund of Peking University.

\section*{Graphical Abstract}

\begin{figure}
	\centering
	\includegraphics[scale=0.42]{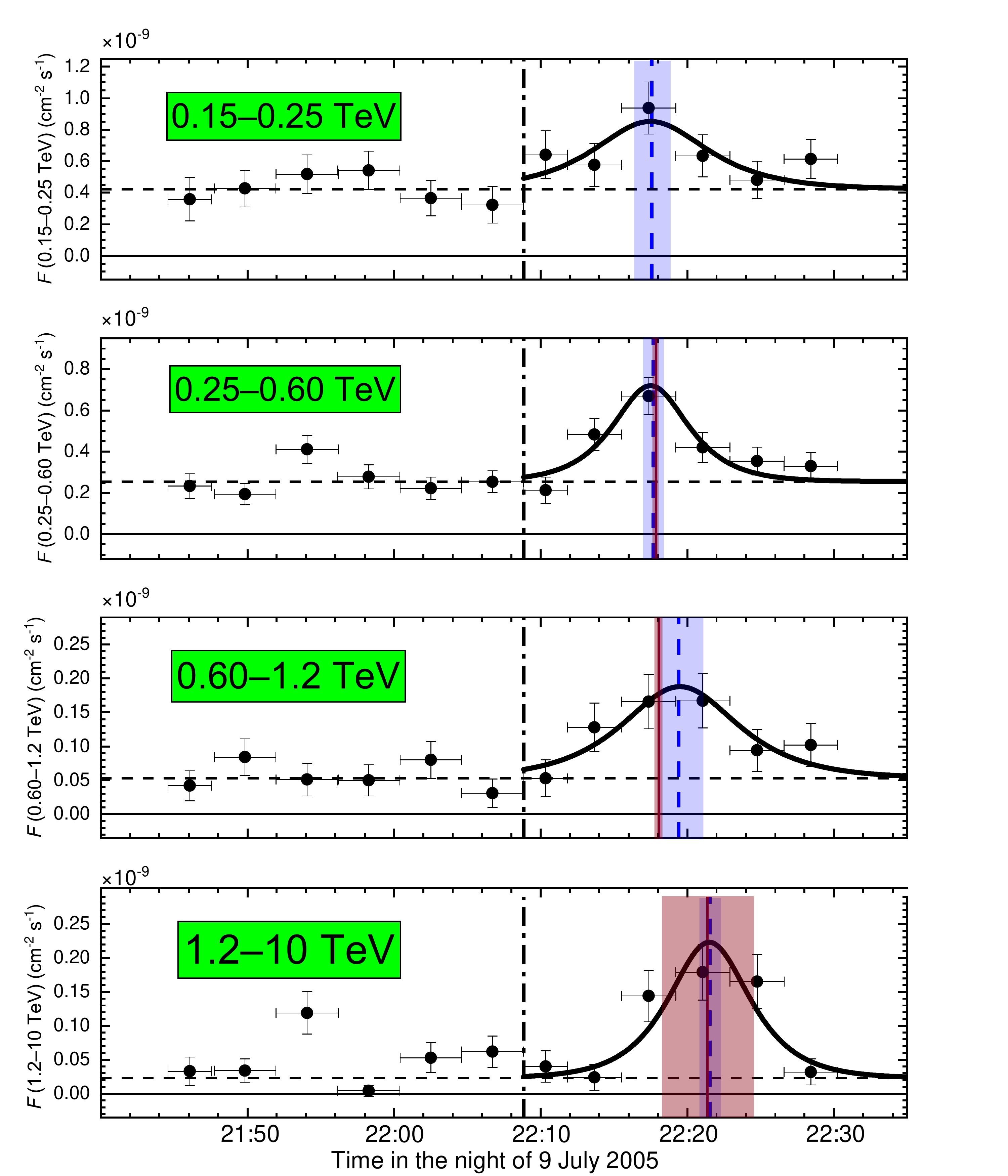}
	\caption{Light curves (LC) for the flare of Markarian 501 in the night on 9 July 2005. The vertical blue dashed lines show the positions of the peaks of these four light curves, and the blue regions show the corresponding errors. The vertical red solid lines show the predicted peak positions determined by light speed variation from GRB photons, and the red regions show the corresponding errors. We immediately find that all predicted time delays are consistent with the observed time delays when errors are taken into account. Therefore time delays for high energy photons from AGNs can be explained by light speed variation determined from GRBs, and we may expect that light speed variation is a general regularity and AGNs can be good candidates for verifying this.}
\end{figure}

Light curves (LC) for the flare of Markarian 501 in the night on 9 July 2005. The vertical blue dashed lines show the positions of the peaks of these four light curves, and the blue regions show the corresponding errors. The vertical red solid lines show the predicted peak positions determined by light speed variation from GRB photons, and the red regions show the corresponding errors. We immediately find that all predicted time delays are consistent with the observed time delays when errors are taken into account. Therefore time delays for high energy photons from AGNs can be explained by light speed variation determined from GRBs, and we may expect that light speed variation is a general regularity and AGNs can be good candidates for verifying this.

\bibliographystyle{unsrt}

\bibliography{ref.bib}

\end{document}